
%


\nopagenumbers
\headline={\ifnum\pageno=1 \hss\thinspace\hss
     \else\hss\folio\hss \fi}
%
\def\heading#1{\vbox to \dimen1 {\vfill}
     \centerline{\bf #1}
     \global\count11=1  
     \vskip \dimen1}
%
\count10 = 0
\def\section#1{\vbox to \dimen2 {\vfill}
    \global\advance\count10 by 1
    \centerline{\tenrm \uppercase\expandafter{\number\count10}.\ {#1}}
    \global\count11=0  
    \vskip \dimen2}
%
\def\subsection#1{\global\advance\count11 by 1
    \vskip \dimen2
    \centerline{{\it {\number\count10.\number\count11}\/}\ {\it #1}}
    \global\count12=0  
    \vskip \dimen2}
%
\def\subsubsection#1{\global\advance\count12 by 1
    \vskip \dimen2
    \centerline{{\tenit {\number\count10.\number\count11.\number\count12}\/}\
{\tenit #1}}
    \vskip \dimen2}
%
%
\def\refindent{\advance\leftskip by 24pt \parindent=-24pt}
%
\def\journal#1#2#3#4#5{{\refindent
                      {#1}        
                      {#2},       
                      {\it #3\/}, 
                      {\bf #4},   
                      {#5}.       
                      \par }}

%
\def\infuture#1#2#3#4{{\refindent
                  {#1}         
                  {#2},        
                  {\it #3\/},  
                  {#4}.        
                   \par }}
%
\def\inbook#1#2#3#4#5#6#7{{\refindent
                         {#1}         
                         {#2},        
                      in {\it #3\/},  
                     ed. {#4}         
                        ({#5}:        
                         {#6}),       
                       p.{#7}.        
                         \par }}
%

%

%
\def\privcom#1#2#3{{\refindent
                  {#1}        
                  {#2},       
                  {#3}.       
                  \par }}
%
\def\phdthesis#1#2#3{{\refindent
                    {#1}                 
                    {#2}, Ph.D. thesis,  
                    {#3}.                
                    \par}}
%
\def\circular#1#2#3#4{{\refindent
                     {#1}          
                     {#2},         
                     {\it #3\/},   
                     {#4}.         
                     \par}}
%
\def\figcap#1#2{{\refindent
                   Fig. {#1}.---   
                        {#2}       
                        \par}}
%

%
\def\eg{{\it e.g.\/\ }}
\def\sun{_\odot}

\def\alwaysmath#1{\ifmmode {#1}
                  \else {$#1\mkern-5mu$} \fi}
\def\teff{\alwaysmath{T_{\rm eff}}}
\def\ni{\noindent}  
\def\msun{\alwaysmath{\rm\,M_\odot}}
\def\sm{\alwaysmath{\rm M_\odot}}

\def\ie{i.e.}
\def\page{\vfill\eject}
\def\etal{{\it et al.}}
\def\lta{\alwaysmath{\; \hbox{\raise 0.6 ex \hbox{$<$}\kern -1.8 ex\lower .5
ex\hbox{$\sim$}}\;}}
\def\gta{\alwaysmath{\;\hbox{\raise 0.6 ex \hbox{$>$}\kern -1.7 ex\lower .5
ex\hbox{$\sim$}}\;}}

\rightline{To Appear in {\it The Astronomical Journal}}

\heading {MASS LOSS IN M67 GIANTS: EVIDENCE FROM ISOCHRONE FITTING}

\vskip 1.5in

\centerline {\bf Michael J. Tripicco}
\centerline {Astronomy Department, University of Maryland}
\centerline {College Park, MD  20742-2421}
\centerline {\bf Ben Dorman}
\centerline {Department of Astronomy, University of Virginia}
\centerline {Charlottesville, VA  22903-0818}
\centerline {\bf and}
\centerline {\bf R. A. Bell}
\centerline {Space Telescope Science Institute}
\centerline {3700 San Martin Drive, Baltimore, MD  21218}
\centerline {(currently on leave from the University of Maryland)}

\page
\heading {ABSTRACT}

We present a study of the stellar content of the open cluster M67.
We have computed new  evolutionary sequences of stellar
models with solar abundance that cover all phases of evolution from the
Zero-Age Main Sequence to the bright end of the Asymptotic Giant Branch
(AGB). The main sequence and giant branch calculations are presented as
isochrones, while the later phases of evolution form a set of
horizontal branch (HB) and AGB tracks of different envelope masses,
\ie\ implicitly assuming various degrees of mass loss on the giant
branch.  The wide agreement on the age (5 Gyr) and metallicity (solar)
of this cluster make it an ideal reference standard, and the generally
excellent fit of our new isochrones to the M67 color-magnitude diagram
(CMD) is consistent with these values.  We examine the fit  between
the calculated and the observed red giant branch (RGB) in particular,
and discuss factors that most influence its quality, {\it viz.} the
mixing-length parameter and the outer boundary conditions used for the
models. The distinct color gap
between the RGB and the clump giants (\ie, the red HB stars) is
compared with the temperature gap between the He-burning tracks
and the computed 5 Gyr RGB (where the mass, M $\approx
1.27$\msun).  This purely differential approach strongly indicates that
the clump giants have M $\lta 0.70\msun\ $, which implies an
amount of mass loss ($\approx 0.6$ \msun) well in excess of that
found in
globular cluster stars. Possible interpretations of this result are
that mass loss in cool low-mass giants increases either with metallicity
or with initial mass. Observational constraints on mass loss
processes tend to favor the former explanation, provided
a mechanism exists that is much more efficient at solar
composition than at the abundances found in metal-rich globular clusters.

\page

\section {INTRODUCTION}

As part of an ongoing project to calculate synthetic integrated spectra
for globular clusters and galaxy nuclei, we have produced a new set of
solar abundance isochrones over the range from 4 to 16 Gyr. The
isochrones have been calculated following VandenBerg (1983) using
similar computational techniques.  The evolutionary tracks have been
continued  up the first-ascent red giant branch  (RGB) to the helium
flash. A number of post-flash tracks have also been computed in a fully
consistent fashion as detailed by Dorman (1992); these comprise the
zero-age Horizontal Branch (ZAHB) as well as the subsequent evolution
to the asymptotic giant branch (AGB).  The (pre-flash) evolutionary
tracks have been transformed into isochrones in a straightforward
manner, and sampled at narrow intervals of temperature and luminosity.
At each of the sampled points we generate a model atmosphere (using
MARCS) and then a synthetic spectrum (via the SSG program). In this way
we can  translate the isochrones directly from the fundamental
\teff\ -- $\log L$ plane to a color-magnitude diagram (CMD).

Proper theoretical modelling of the integrated light from galaxy nuclei
will depend crucially on our ability to compute both high-metallicity
isochrones and stellar spectra over the temperature and gravity range
set by those isochrones.
However, the higher the metallicity the more challenging it becomes to
compute synthetic stellar spectra accurately because of the
great strength of the atomic and molecular features.
For this reason we felt it wise to test the isochrones via
comparison with the CMD of one of the most well-studied old open
clusters in the Galaxy, M67.  The age and metallicity of this cluster
are regarded as well-determined  by recent work using a number of
different techniques.  In addition, its CMD shows an unusually
well-defined RGB and ``clump''.  This enables us to investigate in some
detail one of the more difficult aspects of comparisons
between theory and observation, {\it
viz.} matching the slope of the RGB. Finally, the quality of the data,
coupled with accurate cluster membership information and the new helium
burning tracks, allows us to estimate the masses of the clump giants
and thus deduce the amount of mass lost between the RGB and the
He-burning phases.

A detailed description of the calculation of the evolutionary tracks,
isochrones, model atmospheres and synthetic spectra are presented in
Sec. 2, along with the sources for the M67 observations. In Sec. 3
we show the results of the comparison between
the theoretical and observational data.
Sec. 4 contains a brief discussion of these results,
their uncertainties and the implications for RGB mass loss scenarios.
Our conclusions are summarized in Sec. 5.

\section {THE COLOR-MAGNITUDE DIAGRAM}
\subsection {Theoretical Calculations}

We have calculated solar abundance evolutionary tracks for stars having masses
between 0.4 and 1.4 $M\sun$, using the the evolution code described
in Dorman (1992).
This program is a  rewritten version of University of Victoria code
described by VandenBerg (1983; 1992) which has been augmented by a
detailed treatment of the helium-burning evolutionary phases.
The evolution is
followed from the pre-main sequence Hayashi track to the zero-age main
sequence through central hydrogen exhaustion to the
RGB until termination at the He-flash.
In addition, we have also generated He-burning tracks
for masses between 0.6 and 1.3 $M\sun$.
These begin at the ZAHB (assuming a core mass of 0.469
$M\sun$) and are allowed to evolve up the AGB until thermal pulses begin
to occur, at which point the evolution is terminated.
We adopted a metallicity $Z=0.0169$, a helium fraction $Y=0.27$, and
Los Alamos (Huebner \etal\ 1977) and Alexander (1975; 1981)
low-temperature opacities in these calculations.
We have performed a number of experiments with the new OPAL opacities
(Iglesias, Rogers, and Wilson 1992)
and conclude that their use is unlikely to change the conclusions of
this study, although we consider their possible impact later in this
paper.
The slope of the RGB is, however, significantly affected by the
choice of the surface pressure boundary condition. We have
adopted the semi-empirical surface pressure
grid tables discussed by VandenBerg (1992), which are
based on model atmospheres but include empirical corrections
designed to improve the agreement with derived IR color-temperature relations
for globular clusters.
We also  conducted a number of trials using other choices for the boundary
condition, as we will discuss shortly.
The ratio $\alpha$ of mixing length to pressure scale height that best
fit the M67 RGB sequence was found to be 1.6, slightly larger than
the value used in VandenBerg's recent work.

The evolutionary tracks  (ZAMS through He-flash) have been transformed into
isochrones for ages
between 4 and 16 Gyr using the usual technique of equivalent-evolutionary
points (Prather, 1976; Bergbusch \& VandenBerg, 1992).
Note that the masses of the RGB stars at 4 and 16 Gyr are
$\approx 1.35\msun\  {\rm and } \approx  0.94\msun,\ $ respectively.
Finally, to cover fully the major evolutionary phases found in open and
globular clusters, we add to the isochrones our He-burning
evolutionary tracks, described above.
{\sl We emphasize that
in every sense possible these post-flash
models have been calculated in a fashion consistent with the RGB models.}
This includes matching $\alpha$, the pressure boundary condition,
opacities and so forth.

The next stage in this exercise is to generate synthetic spectra for
stellar models at intervals along the isochrones as well as the
subsequent He-burning evolutionary tracks.  At each selected \teff\ --
$\log g$ point a plane-parallel, flux-constant model atmosphere was
calculated using MARCS (Gustafsson \etal, 1975), followed by a
synthetic spectrum computed using SSG (Bell \& Gustafsson, 1978, 1989;
Gustafsson \& Bell, 1979).  The spacing used between flux points was
0.1 \AA, and all of the spectra were computed between 3000 and 12000
\AA. We point out that for a variety of reasons, models for stars
cooler than about 4000 K are still quite uncertain (see Bell \&
Paltoglou, 1993). One of the main difficulties arises from the
dominance and strong temperature sensitivity of the TiO absorption
bands. Work is in progress to update the handling of molecular species
in the SSG program; for the present paper we have synthesized all of
the spectra without using TiO.  Note that, for reasonable assumptions
about the Initial Mass Function, less than about 7\% of the
total V-band luminosity in a 5 Gyr old solar abundance cluster will arise
from stars cooler than 4000 K.

The last step in the process was to convolve each synthetic spectrum
along the isochrones with filter transmission profiles to determine the
colors.  Though we concentrate here on the broadband colors (using
profiles from Bessell 1990) simulated photometry for other color
systems (\eg DDO, Washington) has also been carried out.  The colors
have been calibrated in a model-independent manner using the Gunn \&
Stryker (1983) spectrophotometric scans as described in Tripicco \&
Bell (1991).  Absolute magnitudes and bolometric corrections were
determined by relating them to our SSG solar model (with $\teff=5770$
K, log g $ =4.44$, $(B-V)=0.64$) and adopting the values $M_{V\sun}=+4.84$ and
B.C.$\sun=-0.12$.

The isochrones themselves will be presented as part of a larger
integrated spectrum array in a forthcoming paper, along with a full set of
calibrated integrated light colors in a variety of bandpasses.

\subsection {M67 Photometry}

The primary source for the M67 photometry used in this paper is the
tabulation given in Mathieu \etal\ (1986). They measured precise radial
velocities for 170 late-type stars in M67, the sample having been selected
on the basis of proper-motion membership probabilities from Sanders
(1977). The photometry itself was drawn from a number of sources,
including Eggen \& Sandage (1964), Racine (1971),
and Janes \& Smith (1984). Table II(b) of the
Mathieu \etal\ paper lists the full set of photometric references.
We have discarded the spectroscopic binaries identified
by Mathieu, Latham \& Griffin (1990).
As the objects of primary concern in this study are the evolved stars, the
sample of proper-motion members  (mostly restricted to stars brighter
than V $=12.8$) provides an excellent basis for comparison with the
theoretical RGB and HB models.
In order to demonstrate the fit of our isochrone to the main-sequence
and fainter turnoff stars we have also taken data directly from Racine (1971).

In addition to the UBV data discussed above, we use K-band magnitudes
for 23 M67 stars published by Cohen, Frogel \& Persson (1978) to determine
accurate temperatures from the $(V-K)-\teff$ relation given in Bell \&
Gustafsson (1989). We use these data to make a limited comparison between
M67 and the isochrones in the fundamental plane, which is presented in the
following section.

\subsection {Properties of M67}

A major aim of the present paper is to confirm that
our newly computed isochrones are accurate over all phases of
stellar evolution, from the unevolved main-sequence to the AGB.
M67 provides an appropriate test case as it has been
extremely well-studied over the years and there seems to be no
major disagreement remaining over its basic properties.
Recent studies concur that the metallicity of M67 is virtually
indistinguishable from the solar value. The status of the cluster as a
fundamental reference standard has
been nicely summarized by Janes (1985); other recent direct
determinations of the metallicity may be found in Hobbs \&
Thorburn (1991) who derived [Fe/H]$=-0.04 \pm 0.12$ from
high-dispersion spectroscopy, and Nissen \etal\ (1987) who
used {\it uvby} photometry to determine a value of
[Fe/H]$=-0.06 \pm 0.07$.

The age of a star cluster cannot, of course, be as directly
determined as the metallicity.
Ages are generally based on isochrone-fitting, which
depends on a number of assumed cluster parameters
such as reddening, distance, helium abundance
as well as on the details of physics underlying the
isochrones themselves.
Thus, it is refreshing to note that virtually all
of the players in the isochrone-fitting game
appear to have reached a consensus
on an age near to 5 Gyr for M67, with uncertainties of only a
fraction of a Gyr remaining.
A detailed discussion is presented by Demarque, Green \& Guenther (1992).
It is worth noting that Hobbs \& Thorburn (1991) use their
high-dispersion spectra
to  determine directly the effective temperature of stars at the turnoff
of M67--their result of $6165 \pm 60$K leads to an age of 5.2 Gyr $\pm 1.0$
based on the \teff(turnoff)-age relationship from our isochrones.
This type of age estimate has the advantage of being independent of
cluster reddening and distance estimates
as well as any color-temperature transformations.
In summary, it appears that adopting an age of 5 Gyr for M67
should satisfy our requirements for carrying out this test of
our newly-calculated isochrones, particularly as our emphasis will
be on the later phases of evolution where age effects are minimal.

\section {RESULTS}
\subsection {Theoretical H-R Diagram}

Figure 1 compares the post-turnoff portion of our 5 Gyr solar abundance
isochrone with a number of evolved M67 stars in the H-R diagram.
Temperatures have been computed from $(V-K)$ colors
as described in the previous section.
The luminosities are based on an apparent distance modulus $(m-M)_V=9.55$
and a functional relationship between bolometric correction and  \teff\
from our set of models. The M67 giants (up to log L/L$\sun = 2.1$) are nicely
matched by the isochrone calculated with VandenBerg's semi-empirical
tables for the
surface pressure boundary condition. Alternative choices for
this boundary condition are also indicated in Figure 1. Using a
scaled-solar T-$\tau$ relation results in a considerably cooler (and
slightly flatter) RGB, partly due to overestimated low-temperature
opacities.
If  used without modification,  the `pure' (MARCS) model atmospheres
produce somewhat bluer giant branches, but
the RGB becomes progressively flatter toward increasing luminosities.
It was precisely this tendency which led VandenBerg (1992)
to modify the MARCS-based pressure boundary conditions
so as to match globular cluster RGBs (see his Fig. 6 and Sec. 4).
The latter are, of course, much older and generally more metal-poor
than the present case.
All of the RGB tracks indicated in Figure 1 were calculated using an
adopted value of 1.6 for $\alpha$; a change to $\alpha=1.5$ results
 in a displacement of the track to the right
in the theoretical HR diagram, with
the RGB becoming nearly 100K cooler at a given luminosity.

\subsection {Color-Magnitude Diagram}

In Figure 2 we compare our solar abundance isochrones with the
color-magnitude diagram of M67. The adopted reddening (E(B-V)=0.032)
matches the value found by Nissen, Twarog, \& Crawford (1987) to
provide the best fit to their {\it uvby}H$\beta$ photometry. (They also
provide a useful discussion of their reddening estimate in the context
of previous results and find it to be consistent.)  Figure 2 shows
our 5 Gyr isochrone to provides quite a good fit to the whole of the
M67 data for an apparent distance modulus of $(m-M)_V=9.55$. The RGB
slope is clearly less than perfect, but the detailed shape is a function
of numerous free parameters. For example,
as outlined in the previous section
the choice of $\alpha$ and the surface pressure boundary condition
strongly affect the position of the RGB.
Fig. 1 suggests that these parameters have been well-chosen
(note, however, that the brightest M67 giants plotted there
are approximately one magnitude fainter than those in Fig. 2).
But transformation to the color-magnitude plane imposes additional
uncertainties.
At a given temperature and surface gravity, the resulting spectrum (and
thus the colors) can be a non-negligible function of the strength of
particular spectral features. This is most significant for the coolest
giant stars, once absorption from the TiO bands begins to dominate (at
temperatures below about 4000 K or $(B-V) \geq 1.4$).
Close to the tip of the RGB, mass loss may also play a role.
Constant-mass giant branches move to the red with decreasing
mass, with the tip cooling by about 100K for each 0.10 \sm\ lost.
This effect will modify the slope of the upper RGB, but will
leave the color $(B-V)_{0,g}$ of the giant branch at the HB level
unchanged.  In the next section we
will argue that a significant fraction of the stellar mass is
evidently lost prior to the stars' arrival on the core-helium burning branch.

\subsection {Clump Giants}

The five M67 clump giants with (V-K)-based temperatures are plotted
along with our He-burning evolutionary tracks in Figure 3. The diagram
is essentially an expanded version  of Fig. 1; the position of the 5 Gyr RGB is
indicated,
as is the ZAHB for masses between 0.60 and 1.30 $M\sun$. The
post-ZAHB tracks are plotted for four representative masses, showing
the manner in which the stars are predicted to rise off the ZAHB,
first cooling slightly and then evolving to higher temperatures.
After approximately 70 Myr of rather steady evolution, the tracks
turn sharply back to the red. The late HB evolution is characterized
by a contracting core and evolution toward the AGB, which is reached
after the core is exhausted and the helium-burning shell has gained
nuclear equilibrium ({\sl cf.} Dorman \etal\ 1993 and references
therein).

Comparison of the position of the M67 clump giants with the theoretical
tracks in Fig. 3 immediately suggests a mass of approximately
$0.70M\sun$ for these stars, compared with $\sim 1.27 \sm$ for
the giants. The straightforward interpretation
is that the stars suffer a larger amount of mass loss before the
helium burning stages than do those in globular clusters.
Basing this on an absolute
comparison (such as is shown in Fig. 3) would not be conclusive,
as the derived temperatures depend sensitively on
such things as the accuracy of the $(V-K)-\teff$ calibration, as well
as the detailed position of the theoretical tracks.
In fact, while
VandenBerg (1985) seems to have suggested
years ago that an unusually
large amount of mass loss may have occured in the M67 clump stars,
the systematic effects referred to above prevented him from
calling this result more than ``a very uncertain prognostication''.
{\it However, because the RGB and HB tracks tend to respond nearly identically
to small changes in these  adopted parameters, the observed color gap
can be used in a purely differential sense to estimate the
clump giant masses.}

Inspection of Fig. 2 shows that the clump in the color-magnitude
diagram of M67 is separated from the RGB by 0.10 mag in
$(B-V)$. This color gap is obviously independent of the choice of
reddening or other cluster parameters and can be accurately measured
because of the clean separation between the clump and RGB, a result
of the richness of M67 and of the membership criteria firmly
established by both proper motion and radial velocity techniques. Next,
relative comparison between model spectra in the proper temperature and
gravity regime indicates that $\Delta(B-V)=0.10$ corresponds to
$\Delta \teff =250$K. This again is quite insensitive to the details
of the temperatures, gravities, or metallicities of the models used.
The effective temperature gap between the 5 Gyr RGB and the He-burning
tracks of various masses (at the same luminosity) have been calculated
and are shown in Table 1. These are tabulated at three representative
points along the tracks: zero-age, \teff -minimum and \teff -maximum.
{\it The differential comparison illustrated by
Table 1 suggests that a temperature gap of 250K can only be explained
if the clump giants have lost
nearly half of their mass before helium burning commences}.
Note that this result
is insensitive to the adopted cluster age, as the RGB shifts quite
slowly with age (about 25K/Gyr in this age range).

\section {DISCUSSION}

The notion that a significant quantity of mass is lost between the red
giant and horizontal branch stages of evolution is  well-established in
stellar evolution theory.
Mass loss must be inferred to explain the color spread in observed
globular cluster CMDs, but it cannot be so large as to inhibit the
helium burning phases of evolution altogether (see, \eg, Renzini 1981).
More direct evidence for mass outflows from cool giants has been
available for some time (see Dupree 1986 and references therein),
from which mass loss rates are computed.
It is difficult or impossible, however, to quantify how much mass is
actually lost during the giant-branch evolution:
in the globular clusters, this can be inferred  from
comparisons between the observed horizontal-branch morphology and
theoretical models.
This is exactly the methodology of the present study.
The metallicity and age of M67 are well-established, and we have shown
that an isochrone calculated for those parameters provides a good fit
to the whole of the cluster CMD.
Recall that the age estimate is grounded not in the
fit to the giant branch in the CMD but to the
main-sequence turnoff region
(as well as in the spectroscopically-derived
temperature of the turnoff).
Thus the RGB mass is thought to be $\approx 1.27\msun$; an initial mass
as low as 1.20 \msun\ (corresponding to a cluster age $\ge 6.2$ Gyr) is
probably ruled out by these constraints.
The mass of the clump giants is somewhat more difficult to establish,
but the purely differential approach adopted here, based on the
excellent data on one hand and a consistent set of evolutionary
computations on the other, points strongly toward a mass of $0.70M\sun$
or less.

Recent observations may actually show the signature of mass outflow in
progress in this cluster. KPNO 4-m echelle spectra at the Ca II
(H \& K) lines of red giants in M67 show asymmetries in the
emission cores that indicate outflow from the brightest stars.
This outflow is clearly present down to $V=9.69$
but appears to vanish by around $V=10$ (Dupree, 1993).
More quantitative conclusions await a full analysis of these new data
but it is extremely suggestive that the mass loss we have proposed here
may in fact be observable as it happens.
Simply pinpointing where on the RGB the signature of mass loss first
appears will provide important constraints on the future modelling
of the underlying mechanism.

The present study may well
represent the first evidence, albeit indirect, for
the amount of mass lost to stars of approximately solar mass and
metallicity, which differ from the stars of Galactic globular clusters
(GCs) both in age and in metallicity.  Our result allows two different
interpretations, \ie\ that the higher degree of mass loss in M67
giants implies (a) a dependence of mass loss on metallicity or (b)
higher degrees of mass loss in more massive stars.  We consider other
evidence and implications of both of these suggestions.

Any change to the
models that tends to increase the intrinsic gap between the
giant branch and the red end of the HB would, of course,
weaken our conclusion.
It is, however, difficult
to destroy it altogether because HB models with $M > 1 \sm$ are also
significantly (0.25$^m$) brighter, with zero-age locations on the
upper branch of the ZAHB sequence illustrated in Figure 1.
Recall that for the mass lost to be similar to that inferred
from globular clusters, the mean mass of the clump stars
should be about 1.1 \sm.
We have experimented with
fitting the observed points with the more massive models
along the upper branch
but in that case one cannot achieve an
acceptable fit to the turnoff region
regardless of the age of the isochrone used.

The HB models themselves are of course
sensitive to changes in composition, input physics and input
parameters, notably the core mass $M_c$. The
composition is, however, well-constrained by observation to be
greater than [Fe/H] $= -0.1$.
We have not attempted a detailed match to isochrones of lower
metallicity. However, models of significantly lower abundance that are
otherwise consistent with the solar metallicity
isochrones (\ie, $\rm [Fe/H] = -0.23$, $\alpha = 1.6$,
$Y = 0.27$) will move the red end of the HB to approximately the
same color as the observed
clump giants.  The corresponding giant branch
will also be bluer, though by somewhat less than
the ZAHB, so that in order
to reproduce the {\sl gap} the data must again be fitted
away from the coolest HB stars, implying $M < 0.9 \sm$.
Thus even for an extreme choice of the
metallicity for this cluster, the mass of the clump giants
would appear to be
significantly lower than the giant branch.

An increase in $M_c$ at the helium flash tends to shift the HB
evolutionary tracks for a given mass blueward.
However at the red end of the sequence
the dominant effect is to raise the luminosity of the models.
Increasing $M_c$ by 0.01 \sm\ shifts our 0.70 \sm\ sequence blueward by
about 20K, and to a luminosity 0.05$^m$ greater. Of course, this would
require slight adjustment to the derived distance modulus and/or the isochrone
fit.  The derived mass of the clump giants would increase slightly;
however we would argue that a much larger change in $M_c$ would be
required to invalidate our conclusions. The new OPAL opacities imply a
slightly higher value for the solar helium abundance and a different
choice for the mixing length. The effect -- a change in Y from 0.27
to about 0.28 -- may produce a
reduction in the core mass (the increased luminosity
resulting from higher helium will be offset by the larger opacity)
which would tend to strengthen the conclusions of this study.  In sum,
the conclusion we have is robust partly because the match to the
isochrones implies that the data are too faint for the
`upper' branch of the ZAHB sequence, and it is
difficult to see how any of the factors that can shift the theoretical models
can alter this fact.

Simulations of GC horizontal-branch (HB) morphology usually invoke
a mass loss formula proposed by Reimers (1975),

$$\dot M = -4 \times 10^{-13} \eta {L\over gR} \space {\rm  \sm\; yr^{-1}},$$

\ni  ($L,\ g\ {\rm and}\ R$  expressed in solar units), in which $\eta$ is a
free parameter whose value is constrained by observation.
This formula was originally derived (with $\eta =1$) from
observations of cool Population I giants using dimensional analysis.
For GC stars with $\rm M < 0.9 \msun$, comparison of the
HB morphology with theoretical tracks implies values of $\eta$ quite
tightly constrained, \ie\  $\eta = 0.4 \pm 0.2$, with the range
generously estimated. The upper bound comes from the
estimate of Renzini (1981) that a value for $\eta$ as large as 0.6 would
suppress helium-burning evolution
altogether, \ie globular clusters would have no horizontal branch stars
at all.
However, integrating the Reimers
formula along our (constant-mass) evolutionary tracks -- which should
give a fair approximation to a self-consistent calculation in which
the mass is actually removed from the model -- yields much smaller
cumulative mass loss than is implied by our differential comparisons.
Importantly, one infers that
more massive stars lose {\it less} mass than the GC stars.
Figure 4 demonstrates these points. We show here the estimated total mass lost
by a model during the ascent to the RGB tip as a function of stellar luminosity
for four different masses: 0.9, 1.2, 1.3, and 2.0 \sm.
We have calculated $\dot M_{\rm Reimers}(L)$ here with the choice $\eta = 1$.
This choice strips sufficient mass from the 0.9 \sm\ sequence --
corresponding roughly to an 18 Gyr isochrone -- that the helium
burning phases would be suppressed.
 However, even with this relatively large value for $\eta$,
the total mass lost from the 1.3\msun\ sequence is
only about 0.3\sm. The 2.0\sm\ track loses much less mass because it reaches
the
helium core flash at significantly lower luminosity, and
is somewhat hotter during its evolution.
Indeed, if the mass loss were included consistently rather than by {\it
post hoc} estimation then there would be even greater difference in the
predicted mass loss as a function of mass.

Of course, Reimers' formula is not the only mass loss `law'
proposed, nor is it based on a physical model of the processes that
result in mass outflows.
Chiosi and Maeder (1986) have tabulated empirical
mass loss formul\ae, and include several different relations for
cool giants.
Dupree (1986) notes that still other empirical formul\ae\ fit the
observational database equally well.
She also provides a useful summary  of the observational constraints upon
mass loss from cool stars.  Those that apply directly to such
`parametrized mass loss' relations are (i) that it
increase with decreasing \teff\ and (ii) that it increase with $L$.
Neither of these will predict a significant increase in mass loss with
mass {\it at fixed composition}; on the contrary, both imply a
variation in the same sense as the Reimers formula.

One obvious inference is that mass loss is a significant function of
composition.  A current idea is that molecular or dust opacity might
allow radiatively driven winds in cool stars in similar fashion to the
UV resonance lines that drive winds from massive stars (see Holzer \&
MacGregor 1985; MacGregor \& Stencel 1992).
The record from the globulars provides no
strong evidence for variation in the masses of HB stars with
metallicity: of course, this issue is  strongly tied to the infamous
`Second Parameter' problem. However, it is possible that the
responsible mechanism may possess a `threshold' effect, in which
mass loss becomes more efficient below a certain \teff.
This seems plausible for a process that involves
grain formation or line-driven winds.
The fact that none of the
observed metal-rich globular clusters show evidence
of strong RGB mass loss (in the form of populations of
RR Lyrae stars or blue HB stars) would imply that the threshold metallicity
lies in the range $-0.5 < [Fe/H] < 0$.

The idea that mass loss is enhanced in metal-rich stars
has some important implications for the interpretation of
galaxy spectra.
In particular, sufficient mass loss from giants in evolved populations
may produce  hot helium-burning stars.
Increasing mass loss with metallicity would enhance this effect,
producing a positive correlation between UV output and metallicity, as
has been observed (see Burstein \etal\ 1988 for observational data; for
an extensive theoretical exploration, see Greggio and Renzini 1990).
On the other hand, if the mechanism for `slow' stellar
winds in red giants indeed operates
less efficiently in metal-poor systems,
then the later stages of evolution for stars in intermediate-age
clusters might be different.
In particular, the core helium burning stars should be more massive.
Since the hydrogen fuel consumption will be larger, they should
produce more massive post-AGB remnants, which implies brighter
planetary nebul\ae.  Finally, the mass of white dwarfs may be somewhat
greater.

The other hypothesis, that mass loss in cool giants actually might
increase with the initial mass of the giant, appears to run
opposite to the observational constraints noted above. If this
were indeed the case, however, it would imply that the
global stellar properties do not predict mass loss rates -
an uncomfortable conclusion for future understanding in this field.
It is, however, intriguing that we may infer that whatever process
is responsible tends to reduce the size of the stellar envelope {\it to} rather
than {\it by} a constant amount of mass. This inference is
of course highly speculative, unless confirmed by further studies on
similar open clusters. In this regard, the mass of
clump giants is perfectly consistent with the data on Pop. I
white dwarf masses, but implies low-mass or non-existent
planetary nebula ejecta from stars with initial mass $\lta 1.3 \sm$.
This cannot be ruled out, but it seems safe to adopt the
more conservative assumption that we have evidence for
increasing mass loss with metallicity.

\section {CONCLUSIONS}

We have performed a comparison between the color-magnitude diagram of
M67 and a new set of theoretical evolutionary models which include all
phases from the unevolved main-sequence through core-helium burning
(horizontal branch) and onto the asymptotic giant branch.
We find that:

\item{(1)} Our 5 Gyr solar abundance isochrone yields an excellent fit to
the whole of the M67 color-magnitude diagram. The model parameters
adopted were $Z=0.0169, Y=0.27,\  {\rm and}\  \alpha=1.6$. Values of
$(m-M)_V=9.55$ and $E(B-V)=0.032$ for M67 provide the best fit.

\item{(2)} A differential technique that compares the gap in color between
clump giants and normal red giants on one hand with the temperature gap
between core He-burning tracks and first-ascent RGB tracks on the other,
strongly indicates that the clump giants
in M67 have masses $\approx 0.70\msun$ or less.
The extremely large
amount of mass loss that we deduce in this study ($\approx 0.6\msun$) is
well in excess of that found for globular cluster stars.

\item{(3)} Possible resolutions of this problem are that degree of mass loss
increases with total stellar mass, or with metallicity. The
observational evidence on mass loss rates appears to favor the
latter choice, since parametrized mass loss formul\ae\ such as the
Reimers `law'
predict a decrease rather than an increase in the mass loss with total
mass. This appears to be the case for any such formula that obeys
the observational constraints on mass loss rates.

\ni In either case, this study represents the first evidence for
enhanced mass loss in solar-type stars. We are investigating other open
clusters in an effort to identify other good examples of this
phenomenon. It seems unlikely, however, that many more cases
as robust as the
present one can be found in which the cluster age and metallicity are as
well determined, and for which such high-quality photometry and
membership data exist.

\vskip 0.2in

We gratefully acknowledge helpful discussions with
Eileen Friel and Bob Rood.
Thanks also to George Paltoglou
for implementing the routine which transforms
evolutionary tracks into isochrones and to
Andrea Dupree for sharing her results in advance
of publication.
This research was supported by NSF
grants AST-9122361, AST-8918461 and NASA grant NAGW-2596.

\heading {REFERENCES}
\journal {Alexander, D. R.}{1975}{ApJS}{29}{363}
\privcom {Alexander, D. R.}{1981}{private communication}
\journal {Bell, R. A. and Gustafsson, B.}{1978}{A\&AS}{34}{229}
\journal {Bell, R. A. and Gustafsson, B.}{1989}{MNRAS}{236}{653}
\infuture{Bell, R. A. and Paltoglou, G.}{1993}{MNRAS}{(in press)}
\journal {Bessell, M.}{1990}{PASP}{102}{1181}
\journal {Bergbusch, P. A. and VandenBerg, D. A.}{1992}{ApJS}{81}{163}
\journal {Burstein, D., Bertola, F., Buson, L.M., Faber, S.M. and Lauer,
T.R.}{1988}{ApJ}{328}{440}
\journal {Chiosi, C. and Maeder, A.}{1986}{ARA\&A}{24}{329}
\journal {Cohen, J. G., Frogel, J. A. and Persson, S. E.}{1978}{ApJ}{222}{165}
\journal {Demarque, P., Green, E. M. and Guenther, D. B.}{1992}{AJ}{103}{151}
\journal {Dorman, B.}{1992}{ApJS}{81}{221}
\infuture{Dorman, B., Rood, R.T., O'Connell, R.W.}{1993}{ApJ}{(submitted)}
\journal {Dupree, A. K.}{1986}{ARA\&A}{24}{377}
\privcom {Dupree, A. K.}{1993}{private communication}
\journal {Eggen, O. J. and Sandage, A. R.}{1964}{ApJ}{140}{130}
\journal {Greggio, L. and Renzini, A.}{1990}{ApJ}{364}{35}
\journal {Gunn, J. E. and Stryker, L. L.}{1983}{ApJS}{52}{121}
\journal {Gustafsson, B. and Bell, R. A.}{1979}{A\&A}{74}{313}
\journal {Gustafsson, B., Bell, R. A., Eriksson, K. and Nordlund,
A.}{1975}{A\&A}{42}{407}
\journal {Hobbs, L. M. and Thorburn, J. A.}{1991}{AJ}{102}{1070}
\inbook {Holzer, T.E. and MacGregor, K.B.}{1985}{Mass Loss from Red Giants} {M.
Morris and B. Zuckerman}{Dordrecht} {Reidel} {229}
\circular {Huebner, W. F., Merts, A. L., Magee, N. H. and Argo, M.F.}{1977}{Los
Alamos Sci. Lab. Rep.}{LA-6760-M}
\journal {Iglesias, C.A., Rogers, F.J. and Wilson, B.G.}{1992}{ApJ}{397}{717}
\inbook  {Janes, K. A.}{1985}{Calibration of Fundamental Stellar
Quantities}{Hayes, D. S., Pasinetti, L. E. and Philip, A. G.
D.}{Dordrecht}{Reidel}{361}
\journal {Janes, K. A. and Smith, G. H.}{1984}{AJ}{89}{487}
\journal {MacGregor, K.B. and Stencel, R.E.} {1992}{ApJ}{397}{644}
\journal {Mathieu, R.D., Latham, D. W., Griffin, R. F. and Gunn, J.
E.}{1986}{AJ}{92}{1100}
\journal {Mathieu, R. D., Latham, D. W. and Griffin, R.
F.}{1990}{AJ}{100}{1859}
\journal {Nissen, P. E., Twarog, B. A. and Crawford, D. L.}{1987}{AJ}{93}{634}
\phdthesis {Prather, M. J.}{1976}{Yale University}
\journal {Racine, R.}{1971}{ApJ}{168}{393}
\journal {Reimers, D.}{1975}{M\'em. Soc. Roy. Sci. Li\`ege, $6^e$ Ser.}{8}{369}
\inbook  {Renzini, A.}{1981}{Effects of Mass Loss on Stellar Evolution}{Chiosi,
C. and Stalio, R.}{Dordrecht}{Reidel}{319}
\journal {Sanders, W. L}{1977}{A\&AS}{27}{89}
\journal {Tripicco, M.J. and Bell, R.A.}{1991}{AJ}{102}{744}
\journal {VandenBerg, D. A.}{1983}{ApJS}{51}{29}
\journal {VandenBerg, D. A.}{1985}{ApJS}{58}{711}
\journal {VandenBerg, D. A.}{1992}{ApJ}{391}{685}


\heading{FIGURE CAPTIONS}

\figcap{1}{
Comparison between our 5 Gyr solar abundance isochrone and
M67 subgiants and giants in the fundamental plane. The parameters
used in computing the isochrones are given in the text. We illustrate
the results
for three different choices for the surface pressure boundary condition
in the calculation of the giant branch evolutionary tracks.
A value of 1.6 for the convective parameter
$\alpha$ was used in all of these cases.
The semi-empirical tables described by VandenBerg (1992) yield
the best fitting models for the evolved stars in M67.
The zero-age horizontal branch for masses $0.6\ {\rm to}\ 1.3 \sm\ $
(also calculated with the semi-empirical boundary condition)
is also indicated.
}
\vskip 1cm

\figcap{2}{
The color-magnitude diagram of M67 is compared with our isochrones.
The apparent distance modulus and reddening adopted for the cluster
are listed. Filled symbols indicate radial velocity members of M67,
while open symbols refer to photometry taken directly
from Racine (1971) for stars
whose membership is based only on proper motion criteria.
The full 5 Gyr solar abundance isochrone (plus zero-age
horizontal branch) is plotted, along with the main-sequence through
turnoff portions of the 4 and 8 Gyr isochrones from the same set.
}

\vfill\eject

\figcap{3}{
Similar to Fig. 1, but concentrating on the He-burning evolutionary phases.
Open circles represent the 5 M67 clump giants with (V-K)-based temperatures.
The zero-age horizontal branch (ZAHB) is plotted as a sequence of
connected filled circles for the range of total masses as indicated.
The evolution from the ZAHB onto the
asymptotic giant branch is indicated by the dashed lines for
$0.6,\ 0.7,\ 0.9\ {\rm and}\ 1.3 \msun$
}
\vskip 1cm

\figcap{4}{
The cumulative mass loss predicted by the Reimers formula with
$\eta=1.0$. Results are plotted for four representative
solar abundance mass tracks.
Note that the total mass lost prior to the onset of He-burning
{\sl decreases} with increasing initial mass and should not exceed
$\approx 0.3\sm$ in a 5 Gyr old cluster (where the giants
begin with masses near 1.27 \msun) even for such a large value
for $\eta$.
}

\bye